\documentclass[preprint,superscriptaddress,nofootinbib]{revtex4}
     
     \usepackage{amssymb,amsmath,graphicx,subfigure,color}
     \usepackage{rotating,multirow,dcolumn}
     \usepackage[colorlinks]{hyperref}
     \allowdisplaybreaks[4]
     \begin{document}

     \title{The nonfactorizable QED correction to the
            $\overline{B}_{s}$ ${\to}$ $D_{s}^{(\ast)}
            {\ell} \bar{\nu}_{\ell}$ decays}
     \author{Yueling Yang}
     \affiliation{Institute of Particle and Nuclear Physics,
                 Henan Normal University, Xinxiang 453007, China}
     \author{Jiazhi Li}
     \affiliation{Institute of Particle and Nuclear Physics,
                 Henan Normal University, Xinxiang 453007, China}
     \author{Liting Wang}
     \affiliation{Institute of Particle and Nuclear Physics,
                 Henan Normal University, Xinxiang 453007, China}
     \author{Junfeng Sun}
     \affiliation{Institute of Particle and Nuclear Physics,
                 Henan Normal University, Xinxiang 453007, China}

     \begin{abstract}
     Considering the nonfactorizable QED corrections, the branching
     ratios and ratios of branching ratios $R(D_{s}^{({\ast})})$
     for the semileptonic $\overline{B}_{s}$ ${\to}$
     $D_{s}^{(\ast)} {\ell} \bar{\nu}_{\ell}$ decays are
     reevaluated.
     It is found that
     (a)
     the QED contributions can enhance the branching ratios and
     reduce the ratios $R(D_{s}^{({\ast})})$.
     (b)
     The $SU(3)$ flavor symmetry holds basically well
     in the ratios $R(D)$-$R(D^{\ast})$ for the semileptonic charmed
     $\overline{B}_{u,d,s}$ decays.
     (c)
     The current theoretical uncertainties of branching ratios
     ${\cal B}(\overline{B}_{s} {\to} D_{s}^{\ast} {\ell} \bar{\nu}_{\ell})$
     from the form factors are very large.

     \href{https://doi.org/10.1140/epjc/s10052-025-14131-y}
          {Eur. Phys. J. C 85, 402 (2025).}

     \href{http://arxiv.org/abs/2502.09883}
          {arXiv:2502.09883.}
     \end{abstract}
     \maketitle

     \section{Introduction}
     \label{sec01}
     The Standard Model (SM) of particle physics is currently the most
     comprehensive theory of the microscopic structure of matter and
     the fundamental interactions.
     The SM has been rigorously validated through numerous
     experiments, and has achieved great success.
     The precision determination of $V_{cb}$ is pivotal to testing
     the Cabibbo–Kobayashi–Maskawa (CKM) pattern for $CP$ violation
     within SM.
     According to the present status of all the existing experiments,
     a high priority of the matrix element $V_{cb}$ extraction
     is from the semileptonic charmed $B$ decays rather
     than the purely leptonic decays, such as $B^{-}_{c}$ ${\to}$
     ${\ell}\bar{\nu}$.
     The semileptonic decays of $B$ mesons to charm have been studied
     extensively by experimentalists and theorists over the years.
     With the increasement of experimental data sample and the
     improvement of measurement precision, the state-of-the-art of
     $V_{cb}$ is ${\vert} V_{cb} {\vert}$ $=$
     $42.2(5){\times}10^{-3}$ and $39.8(6){\times}10^{-3}$ obtained
     respectively from inclusive and exclusive semileptonic
     decays \cite{PhysRevD.110.030001}.
     There is an approximately $3.0 \, {\sigma}$ discrepancies of
     the values.
     In addition, although the ratios of branching ratios,
     \begin{eqnarray}
     R(D) & {\equiv} &
     \frac{ {\cal B}(\overline{B}{\to}D{\tau}^{-}\bar{\nu}_{\tau}) }
          { {\cal B}(\overline{B}{\to}D{\ell}^{-}\bar{\nu}_{\ell}) }
     \label{rd-definition}, \\
     R(D^{\ast}) & {\equiv} &
     \frac{ {\cal B}(\overline{B}{\to}D^{\ast}{\tau}^{-}\bar{\nu}_{\tau}) }
          { {\cal B}(\overline{B}{\to}D^{\ast}{\ell}^{-}\bar{\nu}_{\ell}) }
     \label{rdstar-definition},
     \end{eqnarray}
     with ${\ell}$ $=$ $e$ and ${\mu}$, are free from $V_{cb}$,
     with the average result of the exclusive
     semileptonic $B$ decays provided by the Heavy Flavor Averaging
     Group (HFLAV) \cite{HFLAV},
     there is an approximately $2.2\,{\sigma}$ ($1.9\,{\sigma}$)
     discrepancies between the SM predictions
      $R(D)_{\rm th}$ $=$ $0.296(4)$
     ($R(D^{\ast})_{\rm th}$ $=$ $0.254(5)$)
     and the measurements
      $R(D)_{\rm exp}$ $=$ $0.342(26)$
     ($R(D^{\ast})_{\rm exp}$ $=$ $0.286(12)$);
     if one considers these deviations together with the correlation
     coefficients of $-0.39$, the significance exceeds $3.0\,{\sigma}$.
     This phenomenon has motivated speculation on the lepton
     flavor universality (LFU) within SM.
     It is well-known that
     the isospin symmetry is a good approximation to the branching
     ratios and $R(D^{({\ast})})$ for the exclusive semileptonic
     charmed $B$ decays \cite{PhysRevD.110.030001}.
     The exclusive semileptonic $\overline{B}_{s}$ ${\to}$
     $D_{s}^{({\ast})} {\ell}^{-} \bar{\nu}_{\ell}$ decays
     involving the underlying $b$ ${\to}$ $c$ $+$ $W^{\ast}$
     ${\to}$ $c$ $+$ ${\ell}^{-}$ $+$ $\bar{\nu}_{\ell}$
     weak transition are the
     $U$- or/and $V$-spin cousins of the $\overline{B}$ ${\to}$
     $D^{({\ast})} {\ell}^{-} \bar{\nu}_{\ell}$ decays, will
     inevitably provide some complementary constraints to the
     matrix element $V_{cb}$ and the LFU problem.
     The semileptonic $\overline{B}_{s}$ ${\to}$
     $D_{s}^{({\ast})} {\mu}^{-} \bar{\nu}$ decays are being
     studied extensively by the active LHCb experiment \cite{PhysRevD.101.072004},
     and more semileptonic $\overline{B}_{s}$ ${\to}$
     $D_{s}^{({\ast})} {\ell}^{-} \bar{\nu}_{\ell}$ decays
     will be investigated carefully by the running Belle II experiment
     in the coming years, and by the planning CEPC \cite{CEPC}
     and FCC-ee \cite{FCCee} through the Tara-$Z$ programme
     in the future.

     Recently, we calculated the branching ratios and ratios
     $R(D^{({\ast})})$ for the exclusive semileptonic
     $\overline{B}$ ${\to}$ $D^{({\ast})} {\ell}^{-} \bar{\nu}_{\ell}$
     decays within SM \cite{EPJC.84.1282}.
     In particular, regarding to the open question whether the
     introduction of the novel lepton-flavor-dependent couplings
     beyond SM is necessary to settle the appealing suspicions on LFU,
     we considered the QED nonfactorizable contributions arising
     from the photon exchange interactions between the heavy flavor
     quarks and the charged leptons of different generations.
     It is found that the QED nonfactorizable contributions can affect
     the effective couplings according to the charged lepton flavor,
     the consequent branching ratios and
     ratios $R(D^{({\ast})})$ for the $\overline{B}$ ${\to}$
     $D^{({\ast})} {\ell}^{-} \bar{\nu}_{\ell}$ decays \cite{EPJC.84.1282}.
     In this paper, we will try to extend the one-loop QED corrections
     to the $\overline{B}_{s}$ ${\to}$
     $D_{s}^{({\ast})} {\ell}^{-} \bar{\nu}_{\ell}$ decays,
     further check the practicalities,
     and update theoretical calculations.

     The remaining parts of this paper are as follows.
     The Section \ref{sec02} delineates the theoretical framework for
     the semileptonic $\overline{B}_{s}$ ${\to}$
     $D_{s}^{({\ast})} {\ell}^{-} \bar{\nu}_{\ell}$ decays,
     including the QED corrections to decay amplitudes.
     The numerical results and comments are presented in Section \ref{sec03}.
     The Section \ref{sec04} devotes to a brief summary.
     The form factors and helicity amplitudes are enumerated in
     Appendix \ref{app01} and \ref{app02}.

     \section{Theoretical framework for the $\overline{B}_{s}$ ${\to}$
       $D_{s}^{({\ast})} {\ell}^{-} \bar{\nu}_{\ell}$ decays}
     \label{sec02}

     Within SM, based on the operator product expansion
     technique, the low energy effective Hamiltonian responsible
     for the semileptonic $\overline{B}_{s}$ ${\to}$
     $D_{s}^{({\ast})} {\ell}^{-} \bar{\nu}_{\ell}$ decays is
     written as the product of the $W$-emission actualized quarkic
     and leptonic currents,
    {\em i.e.},
     \begin{eqnarray}
    {\cal H}_{\rm eff} & = &
     \frac{ G_{F} }{ \sqrt{2} } \, V_{cb} \, j_{h,{\mu}} \, j_{\ell}^{\mu}
     \label{eq:Hamiltonian}, \\
     j_{h}^{\mu} & = & \bar{c} \, {\gamma}^{\mu} \, (1-{\gamma}_{5}) \, b
     \label{eq:hadronic-current}, \\
     j_{\ell}^{\mu} & = & \bar{\ell} \, {\gamma}^{\mu} \, (1-{\gamma}_{5}) \, {\nu}_{\ell}
     \label{eq:leptonic-current},
     \end{eqnarray}
     where $G_{F}$ ${\approx}$ $1.166 {\times} 10^{-5}$ ${\rm GeV}^{-2}$
     \cite{PhysRevD.110.030001} is the Fermi coupling constant.

     The decay amplitude is factorized into two parts,
     \begin{equation}
    {\cal A}_{0} \, = \,
    {\langle} \, D_{s}^{({\ast})} \, {\ell}^{-} \, \bar{\nu}_{\ell} \,
    {\vert} \, {\cal H}_{\rm eff} \, {\vert} \, \overline{B}_{s} \, {\rangle}
     \, = \,
     \frac{ G_{F} }{ \sqrt{2} } \, V_{cb} \, H_{\mu} \, L^{\mu}
     \label{eq:amplitude-leading},
     \end{equation}
     where the hadronic and leptonic current matrix elements are
     respectively defined as,
     \begin{eqnarray}
     H_{\mu} & = &
    {\langle} \, D_{s}^{({\ast})} \, {\vert} \, j_{h,{\mu}} \,
    {\vert} \, \overline{B}_{s} \, {\rangle}
     \label{eq:hadronic-current-lorentz}, \\
     L_{\mu} & = &
    {\langle} \, {\ell}^{-} \, \bar{\nu}_{\ell} \, {\vert} \, j_{{\ell},{\mu}} \,
    {\vert} \, 0 \, {\rangle}
     \label{eq:leptonic-current-lorentz}.
     \end{eqnarray}
     Phenomenologically, the leptonic current matrix elements
     $L_{\mu}$ are calculable, and the hadronic current matrix
     elements $H_{\mu}$ are parameterized with the
     $\overline{B}_{s}$ ${\to}$ $D_{s}^{({\ast})}$
     transition form factors.
     In the practical calculation, a useful and common trick is to
     follow the methods described in Ref. \cite{ZPC46.P93}
     and convert the decay amplitudes 
     into helicity representations,
     \begin{eqnarray}
     H_{\mu} \, L^{\mu} & = & g^{{\mu}{\nu}} \, H_{\mu} \,  L_{\nu}
     \nonumber \\ & = &
     \sum\limits_{ {\lambda}, \, {\lambda}^{\prime} }
    {\varepsilon}_{W}^{{\ast}{\mu}}({\lambda}) \,
    {\varepsilon}_{W}^{{\nu}}({\lambda}^{\prime}) \,
     g_{{\lambda},{\lambda}^{\prime}} \, H_{\mu} \,  L_{\nu}
     \nonumber \\ & = & \!\!\!
     \sum\limits_{ {\lambda}, \, {\lambda}^{\prime} }
     \big\{ {\varepsilon}_{W}^{{\ast}{\mu}}({\lambda})\, H_{\mu} \big\}\,
     \big\{ {\varepsilon}_{W}^{{\nu}}({\lambda}^{\prime})\,  L_{\nu} \big\}\,
     g_{{\lambda},{\lambda}^{\prime}}
     \nonumber \\ & = &
     \sum\limits_{ {\lambda}, \, {\lambda}^{\prime} }
     H_{\lambda} \, L_{{\lambda}^{\prime}} \,
     g_{{\lambda},{\lambda}^{\prime}}
     \label{eq:amplitude-helicity},
     \end{eqnarray}
     where ${\varepsilon}_{W}^{{\ast}{\mu}}({\lambda})$ denotes
     the polarization vectors of virtual $W^{\ast}$ boson with
     the helicity components ${\lambda}$ $=$ $+$, $-$, $0$ and $t$.
     The $H_{\lambda}$ $=$ ${\varepsilon}_{W}^{{\ast}{\mu}}({\lambda})\, H_{\mu}$
     and $L_{\lambda}$ $=$ ${\varepsilon}_{W}^{{\nu}}({\lambda})\,  L_{\nu}$
     are respectively called as the hadronic and leptonic helicity
     amplitudes, and they are invariant under the Lorentz
     transformation.
     The helicity amplitudes $H_{\lambda}$ are listed in Appendix
     \ref{app01} and \ref{app02}.

\begin{figure}[h]
     \subfigure[]{ \includegraphics[width=0.25\textwidth,bb=190 630 390 720]{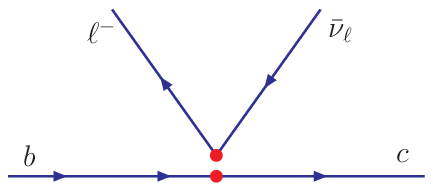} }
     \subfigure[]{ \includegraphics[width=0.25\textwidth,bb=190 630 390 720]{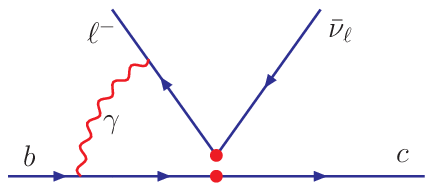} }
     \subfigure[]{ \includegraphics[width=0.25\textwidth,bb=190 630 390 720]{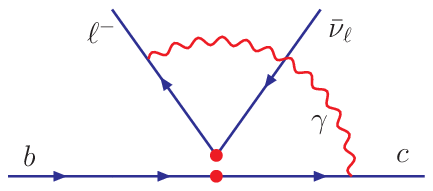} }
     \caption{The Feynman diagrams for the $b$ ${\to}$ $c$ $+$ ${\ell}^{-}$ $+$
          $\bar{\nu}_{\ell}$ decays, where the dots denote the
          local $W$-exchange weak interactions,
          (a) for the leading order contribution,
          (b) and (c) for the QED vertex corrections.}
     \label{fig:vertex}
     \end{figure}

     As the measurements on the semileptonic $B$ weak decays reach
     high precision, it is becoming more and more
     important and necessary to include electroweak radiative
     corrections in comparison of theory and experiment.
     The decay amplitudes are generally rewritten as,
     \begin{equation}
    {\cal A} \, = \, {\cal A}_{0} \, {\eta}_{\rm EW}
     \label{eq:amplitude-nlo},
     \end{equation}
     where ${\cal A}_{0}$ is the leading order amplitudes in
     Eq.(\ref{eq:amplitude-leading}), corresponds to the
     Fig.\ref{fig:vertex} (a).
     The factor ${\eta}_{\rm EW}$ accounts for the short-distance
     electroweak corrections, and has been given in Ref. \cite{NPB.196.83},
     \begin{equation}
    {\eta}_{\rm EW} \, = \, 1 +
     \frac{ 3 \, {\alpha}_{\rm em} }{ 4 \, {\pi} } \,
     ( 1 + 2 \, \bar{Q} ) \, {\ln} \frac{ m_{Z} }{ {\mu} }
     \label{eq:QED-1982},
     \end{equation}
     where the factor proportional to
     $\frac{ 3 \, {\alpha}_{\rm em} }{ 4 \, {\pi} }$
     arises from the lepton self-energy corrections plus
     the photonic corrections to form factor
     plus box diagram contributions involving the
     virtual exchange of $W$ and $Z$ between the hadron and
     lepton, corresponding to Fig.~1 (c), (b) and (a)
     in Ref. \cite{NPB.196.83}, respectively.
     The factor proportional to
     $\frac{ 3 \, {\alpha}_{\rm em} }{ 2 \, {\pi} } \, \bar{Q}$
     arises from box diagram contributions.
     $\bar{Q}$ is the average electric charge of
     the quark doublets,
     $\bar{Q}$ $=$ $ \frac{1}{2} \, ( Q_{b} + Q_{c} ) $ $=$ $\frac{1}{6}$
     for the semileptonic charmed $B$ decays.
     The factor ${\eta}_{\rm EW}$ in Eq.(\ref{eq:QED-1982}) is lepton
     flavor independent, and in most instances
     ${\eta}_{\rm EW}$ ${\approx}$ $1.0066$ \cite{PhysRevD.110.030001}
     with the renormalization scale ${\mu}$ $=$ $m_{B}$ for the
     semileptonic $B$ decays.
     Sometimes an additional overall long-distance factor of
     $1$ $+$ ${\alpha}_{\rm em}\, {\pi}$ arising from Coulomb
     corrections is considered for the neutral $B$ meson decays
     \cite{PhysRevD.41.1736}.
     Aiming to the effective couplings between the gauge bosons and
     the leptons of the LFU problem, we will consider the one-loop
     QED radiative vertex corrections from Fig. \ref{fig:vertex}
     (b) and (c) to the semileptonic $\overline{B}_{s}$ ${\to}$
     $D_{s}^{({\ast})} {\ell}^{-} \bar{\nu}_{\ell}$ decays,
     as argued for the semileptonic $\overline{B}$ ${\to}$
     $D^{({\ast})} {\ell}^{-} \bar{\nu}_{\ell}$ decays in
     Ref. \cite{EPJC.84.1282}.
     In principle, the spectator scattering corrections arising from
     the photon exchange between the spectator $s$ quark and
     the charged lepton should also be taken into account.
     The spectator $s$ quark could be the component of the
     $\overline{B}_{s}$ meson or the $D_{s}^{({\ast})}$ meson.
     The spectator scattering amplitudes will involve the convolution
     integrals of the mesonic wave functions of the $\overline{B}_{s}$
     and $D_{s}^{({\ast})}$ mesons.
     On the one hand, some phenomenological parameters
     may be introduced in the spectator scattering amplitudes,
     as shown in the nonleptonic $B$ decays with the QCD
     factorization approach \cite{NuclPhysB.606.245}.
     On the other hand, theoretical uncertainties from mesonic wave
     functions may overshadow the QED corrections and complicate the
     calculation.
     In order to compare with the electroweak corrections
     in Ref. \cite{NPB.196.83} where the spectator scattering
     corrections are not considered, we will also take the
     spectator scattering corrections out of consideration
     at a first approximation for the time being.
     For the convenience of the following discussion, we
     introduce the symbol of $\tilde{\eta}_{\rm EW}$
     to replace and distinguish from the factor ${\eta}_{\rm EW}$ in
     Eq.(\ref{eq:QED-1982}) blind to the lepton flavors,
     and write the decay amplitudes as
     \begin{equation}
    {\cal A} \, = \, {\cal A}_{0} \, \tilde{\eta}_{\rm EW}
     \, = \, {\cal A}_{0} \, \{ 1 + {\alpha}_{\rm em}\,({\eta}_{b}+{\eta}_{c}) \}
     \label{eq:amplitude-nlo-abc},
     \end{equation}
     where ${\cal A}_{0}$ corresponds to Fig.~\ref{fig:vertex}~(a),
     the factors ${\eta}_{b}$ and ${\eta}_{c}$ stem respectively from
     the QED vector corrections in Fig.~\ref{fig:vertex} (b) and (c).
     The photonic $W$-box diagram of Fig.~1~(a) in Ref. \cite{NPB.196.83}
     within the full SM theory framework
     corresponds to Fig. \ref{fig:vertex} (b) and (c) within the
     effective theory framework.

     After the subtractions of the ultraviolet and infrared divergences,
     the analytic expressions of ${\eta}_{b,c}$ are written as follows
     (see Ref. \cite{EPJC.84.1282} for the more details).
     \begin{eqnarray}
    {\eta}_{b} & = &
     \frac{ Q_{b} \, Q_{\ell} }{ 4\,{\pi} } \, \Big\{
     \Big[ \frac{ t_{b}+s_{b} }{ t_{b}-s_{b} } \,
    {\ln}\Big( \frac{ s_{b} }{ t_{b} } \Big) - 1 \Big] \,
    {\ln}\Big( \frac{ m_{b}^{2} }{ {\mu}_{\overline{\rm MS}}^{2} } \Big)
    +\frac{s_{b}\,{\ln}(s_{b})}{1-s_{b}}
    +\frac{t_{b}\,{\ln}(t_{b})}{1-t_{b}}
     \nonumber \\ &  &
    - \frac{ t_{b}+s_{b} }{ t_{b}-s_{b} }\, \Big[
    2\,{\ln}(t_{b})\,{\ln}\Big(\frac{ t_{b}-s_{b} }{ 1-t_{b} }\Big)
   -2\,{\rm Li}_{2}(t_{b})
   +   {\rm Li}_{2}\Big(\frac{t_{b}}{s_{b}}\Big)
   +i\,{\pi}\,{\ln}\Big(\frac{t_{b}}{s_{b}}\Big)
     \nonumber \\ &  & \qquad\ \quad\,
   -2\,{\ln}(s_{b})\,{\ln}\Big(\frac{ t_{b}-s_{b} }{ 1-s_{b} }\Big)
   +2\,{\rm Li}_{2}(s_{b})
   -   {\rm Li}_{2}\Big(\frac{s_{b}}{t_{b}}\Big) \Big]
   + \frac{ 1 }{ 2 }
     \nonumber \\ &  &
    -\frac{ 1+s_{b} }{ 1-s_{b} }\,{\ln}(s_{b})
    -\frac{ 1+t_{b} }{ 1-t_{b} }\,{\ln}(t_{b}) \Big\}
     \label{vertex-fig02b},
     \end{eqnarray}
     \begin{eqnarray}
    {\eta}_{c} & = &
    -\frac{ Q_{c} \, Q_{\ell} }{ 4\,{\pi} } \, \Big\{
     \Big[ \frac{ t_{c}+s_{c} }{ t_{c}-s_{c} } \,
    {\ln}\Big( \frac{ s_{c} }{ t_{c} } \Big) -4 \Big] \,
    {\ln}\Big( \frac{ m_{c}^{2} }{ {\mu}_{\overline{\rm MS}}^{2} } \Big)
   + \frac{4\,s_{c}\,{\ln}(s_{c})}{1-s_{c}}
   + \frac{4\,t_{c}\,{\ln}(t_{c})}{1-t_{c}}
     \nonumber \\ &  &
   - \frac{ t_{c}+s_{c} }{ t_{c}-s_{c} }\, \Big[
    2\,{\ln}(t_{c})\,{\ln}\Big(\frac{ t_{c}-s_{c} }{ 1-t_{c} }\Big)
   -2\,{\rm Li}_{2}(t_{c})
   +   {\rm Li}_{2}\Big( \frac{ t_{c} }{ s_{c} } \Big)
   +i\,{\pi}\,{\ln}\Big( \frac{ t_{c} }{ s_{c} } \Big)
     \nonumber \\ &  & \qquad\ \quad\,
   -2\,{\ln}(s_{c})\,{\ln}\Big( \frac{t_{c}-s_{c}}{1-s_{c}} \Big)
   +2\,{\rm Li}_{2}(s_{c})
   -   {\rm Li}_{2}\Big(\frac{s_{c}}{t_{c}}\Big)
   -{\ln}\Big( \frac{t_{c}}{s_{c}} \Big) \Big]
     \nonumber \\ &  &
    -\frac{1+s_{c}}{1-s_{c}}\,{\ln}(s_{c})
    -\frac{1+t_{c}}{1-t_{c}}\,{\ln}(t_{c}) + 9 \Big\}
     \label{vertex-fig02c},
     \end{eqnarray}
     where the electric charges $Q_{b}$ $=$ $-1/3$, $Q_{c}$ $=$ $+2/3$
     and $Q_{\ell}$ $=$ $-1$.
     The relations among the kinematic variables are
     \begin{eqnarray}
     m_{b}^{2}\, (s_{b}+t_{b}) & = & +2\, p_{b}\, {\cdot} \, p_{\ell}
     \label{vertex-s+t-b}, \\
     m_{c}^{2}\, (s_{c}+t_{c}) & = & -2\, p_{c}\, {\cdot} \, p_{\ell}
     \label{vertex-s+t-c}, \\
     m_{b}^{2}\, s_{b}\, t_{b} \, = \,
     m_{c}^{2}\, s_{c}\, t_{c} & = & m_{\ell}^{2}
     \label{vertex-st-01}, \\
     2\, p_{b}\, {\cdot} \, p_{\ell} -2\, p_{c}\, {\cdot} \, p_{\ell} & = & q^{2}+ m_{\ell}^{2}
     \label{vertex-ql},
     \end{eqnarray}
     where $m_{b}$, $m_{c}$ and $m_{\ell}$ are respectively the
     mass of the $b$ quark, $c$ quark and the lepton ${\ell}$.
     In the numerical calculation, we will use the approximation
     ${\mu}_{\overline{\rm MS}}$ $=$ $m_{b}$,
     $m_{b}$ ${\approx}$ $m_{B_{s}}$ and
     $m_{c}$ ${\approx}$ $m_{D_{s}^{({\ast})}}$,
     where $m_{B_{s}}$ and $m_{D_{s}^{({\ast})}}$ are
     respectively the mass of the $\overline{B}_{s}$ and
     $D_{s}^{({\ast})}$ mesons.

     It is easily seen from Eq.(\ref{vertex-fig02b}) and Eq.(\ref{vertex-fig02c})
     that ${\eta}_{b}/Q_{b}$ ${\ne}$ ${\eta}_{c}/Q_{c}$ when the
     mass of participating particles is taken into account.
     This differs from the case of Ref. \cite{NPB.196.83}
     where the mass of quarks and leptons participating in the
     decay is small and neglected when compared with
     $m_{W}$ and $m_{Z}$ and the combined electroweak corrections
     associated with the $b$ and $c$ quarks are proportional to
     the term of
     $\frac{ 3 \, {\alpha}_{\rm em} }{ 2 \, {\pi} } \,
     ( Q_{b}\, {\ln} m_{Z}^{2} +  Q_{c}\, {\ln} m_{Z}^{2})$
     $=$ $\frac{ 3 \, {\alpha}_{\rm em} }{ 2 \, {\pi} } \,
     \bar{Q} \, {\ln} m_{Z}$ in Eq.(\ref{eq:QED-1982}).

     The differential decay rate distribution for the
     $\overline{B}_{s}$ ${\to}$ $D_{s}^{({\ast})} {\ell}^{-} \bar{\nu}_{\ell}$
     decays is typically written as \cite{ZPC46.P93},
     \begin{eqnarray}
     \frac{ d\,{\Gamma} }{ d\,q^{2} \ d\,{\cos}{\theta} }
     & = & {\vert} {\eta}_{\rm EW} {\vert}^{2}\,
     \frac{ G_{F}^{2} \, {\vert} V_{cb} {\vert}^{2} \,
            {\vert} \vec{p} \, {\vert} \, q^{2} }
          { 256 \, {\pi}^{3} \, m_{B_{s}}^{2} } \,
     \Big( 1 - \frac{ m_{\ell}^{2} }{ q^{2} } \Big)^{2}
     \nonumber \\ &   &
     \big\{ \big[ H_{U} \, ( 1 + {\cos}^{2}{\theta} )
         + 2 \, H_{L} \, {\sin}^{2}{\theta}
         + 2 \, H_{P} \, {\cos}{\theta} \big]
     \nonumber \\ &  & \!\!\!\! \!\!\!\! \!\!\!
     + \frac{ m_{\ell}^{2} }{ q^{2} } \, \big[ 2 \, H_{S}
          + 2 \, H_{L} \, {\cos}^{2}{\theta}
          + 4 \, H_{SL} \, {\cos}{\theta}
          + H_{U} \, {\sin}^{2}{\theta} \big] \big\}
     \label{eq:differential-width-dq2-dcos},
     \end{eqnarray}
     where ${\vert} \vec{p} \, {\vert}$ is the momentum of the $D_{s}^{({\ast})}$
     meson in the rest frame of the $\overline{B}_{s}$ meson.
     $q$ is the momentum of virtual $W^{\ast}$ boson,
     $q$ $=$ $p_{B_{s}}$ $-$ $p_{D_{s}^{({\ast})}}$ $=$
     $p_{\ell}$ $+$ $p_{\bar{\nu}}$.
     ${\theta}$ denotes the polar angular between the
     $D_{s}^{({\ast})}$ meson and the lepton ${\ell}^{-}$.
     \begin{eqnarray}
     H_{U} & = & {\vert} H_{+} {\vert}^{2}  + {\vert} H_{-} {\vert}^{2}
     \label{eq:unpolarized-transverse-component}, \\
     H_{P} & = & {\vert} H_{+} {\vert}^{2}  - {\vert} H_{-} {\vert}^{2}
     \label{eq:parity-odd-component}, \\
     H_{L} & = & {\vert} H_{0} {\vert}^{2}
     \label{eq:longitudinal-component}, \\
     H_{S} & = & {\vert} H_{t} {\vert}^{2}
     \label{eq:scalar-component}, \\
     H_{SL} & = & {\rm Re}(H_{t}\,H_{0}^{\ast})
     \label{eq:scalar-longitudinal-interference-component},
     \end{eqnarray}
     denote respectively the unpolarized-transverse,
     parity-odd, longitudinal, scalar, scalar-longitudinal
     interference components of the hadronic amplitudes.
     $H_{\pm}$, $H_{0}$ and $H_{t}$ are the helicity amplitude
     in Eq.(\ref{eq:amplitude-helicity}), and displayed in Appendix
     \ref{app01} and \ref{app02}.

     \section{Numerical results and discussion}
     \label{sec03}
     It is easily seen from Eq.(\ref{vertex-fig02b}) and Eq.(\ref{vertex-fig02c})
     that the factor $\tilde{\eta}_{\rm EW}$ is a
     function of variable $q^{2}$, ${\cos}{\theta}$, and the mass
     of lepton $m_{\ell}$, and very different from the
     lepton-flavor-universal factor ${\eta}_{\rm EW}$ of
     Eq.(\ref{eq:QED-1982}), as discussed in Ref. \cite{EPJC.84.1282}
     for the semileptonic $\overline{B}$ ${\to}$
     $D^{({\ast})} {\ell}^{-} \bar{\nu}_{\ell}$ decays.
     This implicitly indicates the nonfactorizable corrections to the
     effective couplings may provide a possible solution/scheme to
     the LFU problem, even without the introduction of some irregular
     couplings beyond SM.
     To provide a quantitative impression of the QED effects on the
     $\overline{B}_{s}$ ${\to}$ $D_{s}^{({\ast})} {\ell}^{-} \bar{\nu}_{\ell}$
     decays, with the
     input parameters listed in Table \ref{tab:input}
     and the form factors from lattice QCD
     \cite{PhysRevD.101.074513,PhysRevD.105.094506}
     illustrated in Appendix \ref{app01} and \ref{app02},
     the numerical results on the branching ratios and ratios of branching ratios
     are respectively presented in Table \ref{tab:br} and \ref{tab:rDs}.
     It is seen from Table \ref{tab:input} that the current measurement
     precision of the particle mass is very high.
     Taking the $\overline{B}_{s}$ ${\to}$ $D_{s}^{\ast} {\tau}^{-} \bar{\nu}_{\tau}$
     decay as an example, our study shows that the relative error
     of branching ratio (and $R(D_{s}^{\ast})_{\ell}$) from the
     particle mass is about $0.1\%$ (and $0.04\%$).
     The relative errors of branching ratios for all the
     $\overline{B}_{s}$ ${\to}$ $D_{s}^{(\ast)} {\ell}^{-} \bar{\nu}_{\ell}$
     decays from ${\tau}_{B_{s}}$ and ${\vert} \, V_{cb} \, {\vert}$
     are respectively about $0.7\%$ and $3.0\%$.
     The ratios of $R(D_{s})$ and $R(D_{s}^{\ast})$ have nothing to
     do with ${\tau}_{B_{s}}$ and ${\vert} \, V_{cb} \, {\vert}$.
     The shape lines of form factors versus $q^{2}$, especially for
     the $\overline{B}_{s}$ ${\to}$ $D_{s}^{(\ast)}$ transitions
     in Fig. \ref{fig:ffP2V}, are not well determined yet.
     The main theoretical uncertainties come from the form factors.
     There are some comments on the numerical results.

     \begin{table}[h]
     \caption{Values of input parameters given by PDG \cite{PhysRevD.110.030001},
              where their central values are regarded as the default inputs
              unless otherwise specified.}
     \label{tab:input}
     \begin{ruledtabular}
     \begin{tabular}{llll}
       $ m_{B_{s}} $        $=$ $ 5366.93(10) $ MeV,
     & $ m_{D_{s}} $        $=$ $ 1968.35(7)  $ MeV,
     & $ {\tau}_{B_{s}} $   $=$ $ 1527(11)    $ fs,
     & $ m_{\mu}   $        $=$ $ 105.658     $ MeV, \\
       $ {\vert} \, V_{cb} \, {\vert} $  $=$  $ 39.8(6) \, {\times} \, 10^{-3}$,
     & $ m_{D_{s}^{\ast}} $ $=$ $ 2112.2(4)   $ MeV,
     & $ m_{e}     $        $=$ $ 0.511       $ MeV,
     & $ m_{\tau}  $        $=$ $ 1776.93(9)  $ MeV.
     \end{tabular}
     \end{ruledtabular}
     \end{table}
     \begin{table}[h]
     \caption{Branching ratios for the semileptonic $\overline{B}_{s}$
     ${\to}$ $D_{s}^{(\ast)} {\ell}^{-} \bar{\nu}_{\ell}$ decays in
     the unit of percentage, where the theoretical uncertainties
     come only from form factors.}
     \label{tab:br}
     \begin{ruledtabular}
     \begin{tabular}{lcccc}
     \multicolumn{1}{c}{modes}
     & ${\eta}_{\rm EW}$ $=$ $1.0066$
     & $\tilde{\eta}_{\rm EW}$ (this work)
     & PDG \cite{PhysRevD.110.030001}
     & LHCb \cite{PhysRevD.101.072004}
       \\ \hline
       $D_{s} e^{-} \bar{\nu}_{e}$
     & $ 2.23 \, {\pm} \, 0.12 $
     & $ 2.77 \, {\pm} \, 0.15 $
     & ---
     & ---
       \\
       $D_{s} {\mu}^{-} \bar{\nu}_{\mu}$
     & $ 2.22 \, {\pm} \, 0.12 $
     & $ 2.31 ^{+0.13}_{-0.12} $
     & $ 2.31 \, {\pm} \, 0.21 $
     & $ 2.49 \, {\pm} \, 0.24 $
       \\
       $D_{s} {\tau}^{-} \bar{\nu}_{\tau}$
     & $ 0.66 \, {\pm} \, 0.04 $
     & $ 0.67 \, {\pm} \, 0.04 $
     & ---
     & ---
       \\ \hline
       $D_{s}^{\ast} e^{-} \bar{\nu}_{e}$
     & $ 5.09 ^{+2.24}_{-1.64} $
     & $ 6.33 ^{+2.79}_{-2.04} $
     & ---
     & ---
       \\
       $D_{s}^{\ast} {\mu}^{-} \bar{\nu}_{\mu}$
     & $ 5.06 ^{+2.20}_{-1.62} $
     & $ 5.28 ^{+2.30}_{-1.69} $
     & $ 5.2 \, {\pm} \, 0.5 $
     & $ 5.38 \, {\pm} \, 0.60$
       \\
       $D_{s}^{\ast} {\tau}^{-} \bar{\nu}_{\tau}$
     & $ 1.26 ^{+0.26}_{-0.23} $
     & $ 1.26 ^{+0.26}_{-0.23}$
     & ---
     & ---
     \end{tabular}
     \end{ruledtabular}
     \end{table}
     \begin{table}[h]
     \caption{Ratios of branching ratios for the semileptonic
     $\overline{B}_{s}$ ${\to}$ $D_{s}^{(\ast)} {\ell}^{-} \bar{\nu}_{\ell}$
     decays, where the theoretical uncertainties come only from form factors.}
     \label{tab:rDs}
     \begin{ruledtabular}
     \begin{tabular}{cccc}
       Ratios
     & ${\eta}_{\rm EW}$ $=$ $1.0066$
     & $\tilde{\eta}_{\rm EW}$ (this work)
     & LHCb
       \\ \hline
       $ R(D_{s})_{e} $
     & $ 0.298 ^{+0.019}_{-0.016} $
     & $ 0.240 ^{+0.015}_{-0.013} $
     & ---
       \\
       $ R(D_{s})_{\mu} $
     & $ 0.299 ^{+0.018}_{-0.016} $
     & $ 0.288 ^{+0.017}_{-0.016} $
     & ---
       \\
       $ R(D_{s})_{\ell}$
     & $ 0.299 ^{+0.018}_{-0.016} $
     & $ 0.262 ^{+0.016}_{-0.015} $
     & ---
       \\ \hline
       $ R(D_{s}^{\ast})_{e} $
     & $ 0.248 ^{+0.063}_{-0.046} $
     & $ 0.199 ^{+0.051}_{-0.037} $
     & ---
       \\
       $ R(D_{s}^{\ast})_{\mu} $
     & $ 0.249 ^{+0.061}_{-0.045} $
     & $ 0.239 ^{+0.058}_{-0.043} $
     & $ 0.249 $ \cite{NCC.45.120}
       \\
       $ R(D_{s}^{\ast})_{\ell} $
     & $ 0.248 ^{+0.062}_{-0.045} $
     & $ 0.217 ^{+0.054}_{-0.040} $
     & ---
       \\ \hline
       $ \frac{ {\cal B}(D_{s}        e^{-} \bar{\nu}_{e}) }
              { {\cal B}(D_{s}^{\ast} e^{-} \bar{\nu}_{e}) } $
     & $ 0.438 ^{+0.243}_{-0.150} $
     & $ 0.438 ^{+0.243}_{-0.150} $
       \\
       $ \frac{ {\cal B}(D_{s}        {\mu}^{-} \bar{\nu}_{\mu}) }
              { {\cal B}(D_{s}^{\ast} {\mu}^{-} \bar{\nu}_{\mu}) } $
     & $ 0.438 ^{+0.241}_{-0.149} $
     & $ 0.438 ^{+0.241}_{-0.149} $
     & $ 0.464 \, {\pm} \, 0.045 $ \cite{PhysRevD.101.072004}
       \\
       $ \frac{ {\cal B}(D_{s}        {\tau}^{-} \bar{\nu}_{\tau}) }
              { {\cal B}(D_{s}^{\ast} {\tau}^{-} \bar{\nu}_{\tau}) } $
     & $ 0.527 ^{+0.155}_{-0.116} $
     & $ 0.527 ^{+0.155}_{-0.116} $
     \end{tabular}
     \end{ruledtabular}
     \end{table}
     \begin{table}[h]
     \caption{Contributions of transverse, longitudinal and scalar
       helicity amplitudes for the
       $\overline{B}_{s}$ ${\to}$ $D_{s}^{\ast} {\ell}^{-} \bar{\nu}_{\ell}$
       decays (in the unit of percentage), where the fractions
       $f_{\perp}$ $=$ ${\Gamma}_{U}/{\Gamma}$,
       $f_{L}$ $=$ ${\Gamma}_{L}/{\Gamma}$,
       $f_{S}$ $=$ ${\Gamma}_{S}/{\Gamma}$, and partial decay width
       ${\Gamma}_{i}$ corresponds to the $H_{i}$ with $i$ $=$ $U$, $L$, $S$
       in Eq.(\ref{eq:unpolarized-transverse-component}),
       Eq.(\ref{eq:longitudinal-component}),
       Eq.(\ref{eq:scalar-component}), respectively.}
     \label{tab:helicity}
     \begin{ruledtabular}
     \begin{tabular}{cccc}
       case
     & $f_{\perp}$
     & $f_{L}$
     & $f_{S}$   \\ \hline
       ${\ell}$ $=$ $e$      & $49.9$ & $50.1$ & ${\sim}$ $0$ \\
       ${\ell}$ $=$ ${\mu}$  & $49.9$ & $49.7$ & $0.4$ \\
       ${\ell}$ $=$ ${\tau}$ & $56.0$ & $36.4$ & $7.6$
     \end{tabular}
     \end{ruledtabular}
     \end{table}

     (1)
     Theoretically, the underlying dynamic mechanism is the same for the
     $\overline{B}_{s}$ ${\to}$ $D_{s} {\ell}^{-} \bar{\nu}_{\ell}$
     (or $\overline{B}_{s}$ ${\to}$ $D_{s}^{\ast} {\ell}^{-} \bar{\nu}_{\ell}$)
     decays with different final leptons.
     The partial decay width is proportional to the volume size of phase space.
     As the mass of the charged lepton increases,
     the corresponding phase space becomes more compacted
     due to the energy and momentum conservation,
     and branching ratio also decreases accordingly, {\it i.e.},
     ${\cal B}(\overline{B}_{s} {\to} D_{s} e^{-} \bar{\nu}_{e})$ ${\ge}$
     ${\cal B}(\overline{B}_{s} {\to} D_{s} {\mu}^{-} \bar{\nu}_{\mu})$ $>$
     ${\cal B}(\overline{B}_{s} {\to} D_{s} {\tau}^{-} \bar{\nu}_{\tau})$
     and
     ${\cal B}(\overline{B}_{s} {\to} D_{s}^{\ast} e^{-} \bar{\nu}_{e})$ ${\ge}$
     ${\cal B}(\overline{B}_{s} {\to} D_{s}^{\ast} {\mu}^{-} \bar{\nu}_{\mu})$ $>$
     ${\cal B}(\overline{B}_{s} {\to} D_{s}^{\ast} {\tau}^{-} \bar{\nu}_{\tau})$
     with either the constant ${\eta}_{\rm EW}$ or the lepton-flavor-dependent
     $\tilde{\eta}_{\rm EW}$
     in Table \ref{tab:br}, which further leads to the relationship
     $R(D_{s})_{e}$ ${\le}$ $R(D_{s})_{\mu}$
     and
     $R(D_{s}^{\ast})_{e}$ ${\le}$ $R(D_{s}^{\ast})_{\mu}$
     in Table \ref{tab:rDs},
     similarly to cases for the semileptonic charmed
     $B$ decays \cite{EPJC.84.1282}.

     (2)
     In Eq.(\ref{vertex-fig02b}), the term
     ${\ln}(s_{b}/t_{b})$ $=$ ${\ln}(s_{b}t_{b}/t_{b}^{2})$
     ${\propto}$ ${\ln}(m_{\ell}^{2}/m_{b}^{2})$.
     Similarly, in Eq.(\ref{vertex-fig02c}), the term
     ${\ln}(s_{c}/t_{c})$ ${\propto}$ ${\ln}(m_{\ell}^{2}/m_{c}^{2})$.
     The electromagnetic correction factors ${\eta}_{b,c}$ are
     closely related to the charged lepton mass.
     It is easily seen from Table \ref{tab:br} that
     branching ratios with $\tilde{\eta}_{\rm EW}$ are larger than
     those with ${\eta}_{\rm EW}$.
     The nonfactorizable QED contributions to branching ratio
     for the semitauonic decays are indistinguishable,
     because the lepton ${\tau}$ is massive.
     This leads to the ratios $R(D_{s}^{({\ast})})$ with
     $\tilde{\eta}_{\rm EW}$ generally less than the corresponding
     ones with ${\eta}_{\rm EW}$ in Table \ref{tab:rDs}.
     Here, it should be pointed out that
     $R(D_{s}^{\ast})_{\mu}$ $=$ $0.249$ given by Ref. \cite{NCC.45.120}
     is just an estimated value based on the preliminary LHCb analysis
     of signal, normalization and backgrounds.
     The measured branching ratios for the $\overline{B}_{s}$ ${\to}$
     $D_{s}^{(\ast)} {\tau}^{-} \bar{\nu}_{\tau}$ decays
     are still not available.
     The expected values of $R(D_{s}^{(\ast)})_{\mu}$ with
     $\tilde{\eta}_{\rm EW}$ are basically in accord with
     those with ${\eta}_{\rm EW}$ within theoretical uncertainties
     in Table \ref{tab:rDs}.
     What's more, it is worth noting that branching ratios for the
     semimuonic $\overline{B}_{s}$ decays with $\tilde{\eta}_{\rm EW}$ seem
     to be in better agreement with the available data
     \cite{PhysRevD.110.030001,PhysRevD.101.072004},
     although both the theoretical and experimental uncertainties
     are still large.
     Branching ratios for the semielectronic and semitauonic decays
     in Table \ref{tab:br} provide a ready and helpful reference
     for the future experimental measurements.

     \begin{figure}[h]
     \includegraphics[width=0.3\textwidth]{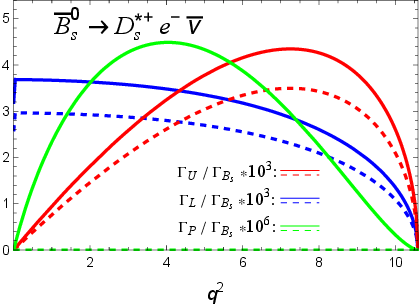} \quad
     \includegraphics[width=0.3\textwidth]{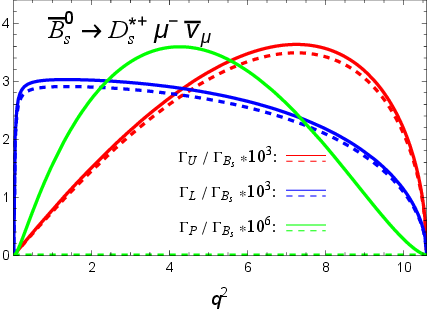} \quad
     \includegraphics[width=0.3\textwidth]{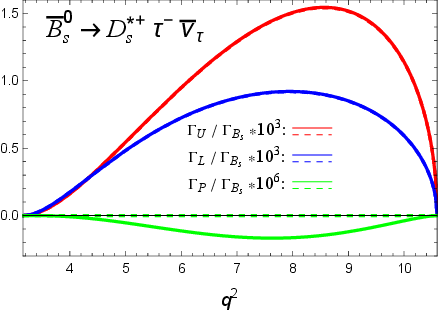} \\ ~ \\
     \includegraphics[width=0.3\textwidth]{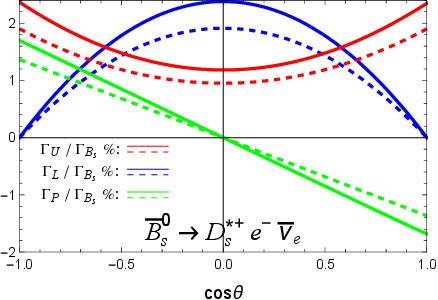} \quad
     \includegraphics[width=0.3\textwidth]{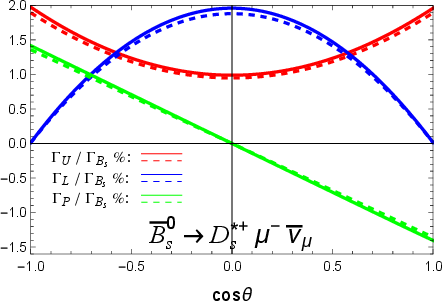} \quad
     \includegraphics[width=0.3\textwidth]{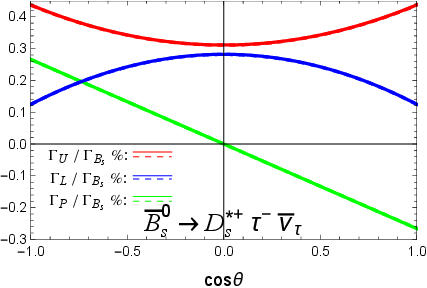}
     \caption{Contributions of different helicity amplitudes for the
       $\overline{B}_{s}$ ${\to}$ $D_{s}^{\ast} {\ell}^{-} \bar{\nu}_{\ell}$
       decays, where the solid (dashed) lines correspond to the
       $\tilde{\eta}_{\rm EW}$ (${\eta}_{\rm EW}$) case.}
     \label{fig:helicity}
     \end{figure}
     \begin{figure}[h]
     \includegraphics[width=0.4\textwidth]{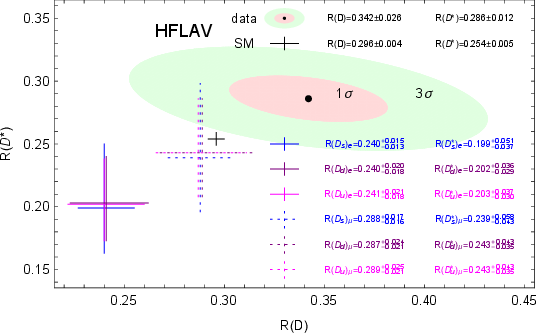}
     \caption{The correlation distribution of ratios
     $R(D)$-$R(D^{\ast})$ for the $\overline{B}_{q}$
     ${\to}$ $D_{q}^{({\ast})} {\ell}^{-} \bar{\nu}_{\ell}$
     decays, where the theoretical values of
     $R(D_{u,d}^{({\ast})})_{\ell}$ are from Ref. \cite{EPJC.84.1282},
     and HFLAV results from Ref. \cite{HFLAV}.}
     \label{fig:rd}
     \end{figure}
     \begin{figure}[h]
     \includegraphics[width=0.3\textwidth]{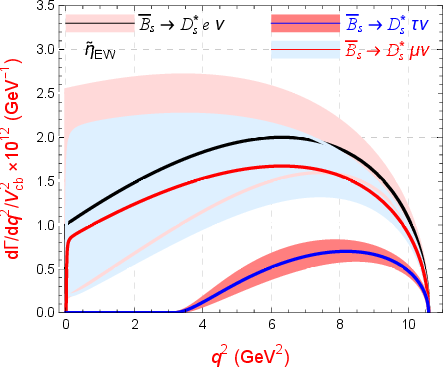} \quad
     \includegraphics[width=0.3\textwidth]{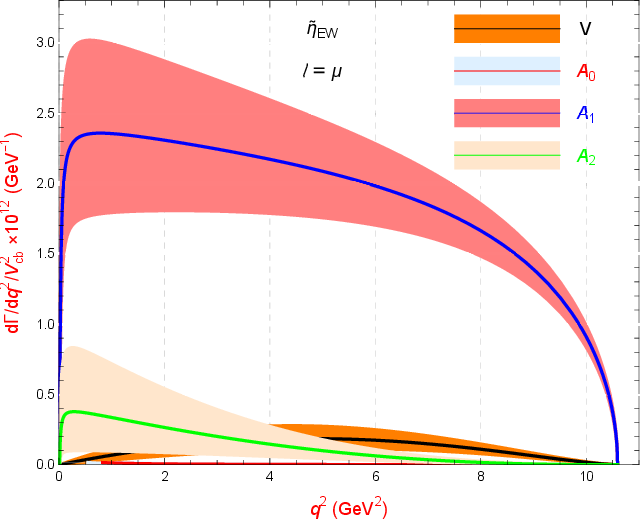} \quad
     \includegraphics[width=0.3\textwidth]{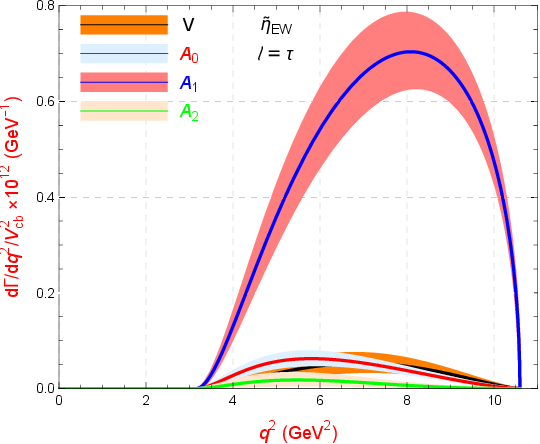}
     \caption{The differential decay rate distributions for the
     $\overline{B}_{s}$ ${\to}$ $D_{s}^{\ast} {\ell}^{-} \bar{\nu}_{\ell}$
     decays.}
     \label{fig:wq2}
     \end{figure}

     (3)
     With either ${\eta}_{\rm EW}$ or $\tilde{\eta}_{\rm EW}$,
     branching ratios for $\overline{B}_{s}$
     ${\to}$ $D_{s}^{\ast} {\ell}^{-} \bar{\nu}_{\ell}$ decays
     are about twice as much as those for $\overline{B}_{s}$ ${\to}$
     $D_{s} {\ell}^{-} \bar{\nu}_{\ell}$ decays with the same
     final leptons, which indicates the significant role of
     the transverse helicity amplitudes.
     The contributions of different helicity amplitudes are
     shown in Table \ref{tab:helicity}
     and Fig. \ref{fig:helicity}.
     It is seen that
     (a)
     the nonfactorizable QED corrections enhance simultaneously
     both the transverse and longitudinal amplitudes depending on
     the charged lepton mass.
     The lighter the charged lepton, the more obvious the enhancement.
     For the semitauonic decay, the enhancement is almost imperceptible.
     (b)
     The transverse (longitudinal) fractions $f_{\perp}$ ($f_{L}$)
     increases (decreases) with the charged lepton mass.
     In addition, $f_{\perp}$ exceeds gradually $f_{L}$ with the
     increase of $q^{2}$, as the distributions of helicity amplitudes
     $H_{\pm}$ and $H_{0}$ in Fig. \ref{fig:ffP2V}.
     (c)
     Although the magnitudes of helicity amplitudes $H_{0}$ and $H_{t}$
     are competitive at the small $q^{2}$ regions in Fig. \ref{fig:ffP2P}
     and \ref{fig:ffP2V}, the contribution of $H_{S}$ is strongly
     suppressed by $m_{\ell}^{2}$ in comparison with those of $H_{L}$
     in Eq.(\ref{eq:differential-width-dq2-dcos}), which leads to
     the relative smaller fraction $f_{S}$ and increasing $f_{S}$
     with the charged lepton mass.
     (d) It is seen from Fig. \ref{fig:helicity} that in the regions
     of ${\cos}{\theta}$ ${\in}$ $[-1,0]$ or
     ${\cos}{\theta}$ ${\in}$ $[0,+1]$,
     the relative fractions of transverse ${\Gamma}_{U}$,
     longitudinal ${\Gamma}_{L}$, and parity-odd ${\Gamma}_{P}$
     contributions are ‌comparable in size.
     The distributions of the transverse ${\Gamma}_{U}$ and
     longitudinal ${\Gamma}_{L}$ contributions are basically
     symmetric with respect to ${\cos}{\theta}$ from $-1$ to $+1$.
     The distributions of parity-odd ${\Gamma}_{P}$ contributions
     are basically antisymmetric with respect to ${\cos}{\theta}$,
     which results in the total parity-odd ${\Gamma}_{P}$
     contributions very small with $\tilde{\eta}_{\rm EW}$
     and zero with ${\eta}_{\rm EW}$.

     (4)
     Theoretically, the $SU(3)$ flavor symmetry holds basically well
     in the ratios $R(D)$-$R(D^{\ast})$ for the semileptonic
     $\overline{B}_{u,d,s}$ decays, see Fig. \ref{fig:rd}.
     It is expected that the precise measurement of the
     semileptonic $\overline{B}_{s}$ ${\to}$
     $D_{s}^{({\ast})} {\ell}^{-} \bar{\nu}_{\ell}$ decays in the
     future experiments
     will provide valuable constraints and helpful information on the
     prominent CKM element $V_{cb}$ and the interesting LFU problem
     shown up in the semileptonic charmed $\overline{B}$ decays.

     (5)
     The theoretical uncertainties of branching ratios for the
     $\overline{B}_{s}$ ${\to}$ $D_{s}^{\ast} {\ell} \bar{\nu}_{\ell}$
     decays from the form factors are very large for the moment,
     especially for the ${\ell}$ $=$ $e$ and ${\mu}$ cases,
     which make the extraction of the CKM element $V_{cb}$ and
     the investigation of nonfactorizable QED effects on the LFU
     virtually impossible.
     It is clearly seen from Fig. \ref{fig:wq2} that
     (a)
     the dominant contributions to the decay width are
     from the form factor $A_{1}$.
     (b)
     The form factor $A_{0}$ contributes to only the helicity
     amplitude $H_{t}$ in Eq.(\ref{eq:ht-Bs2Dsv}), and the scalar
     hadronic amplitudes are strongly suppressed by $m_{\ell}^{2}$
     in Eq.(\ref{eq:differential-width-dq2-dcos}).
     So the contributions from $A_{0}$ to the decay width
     are negligibly small for the $\overline{B}_{s}$ ${\to}$
     $D_{s}^{\ast} e^{-} \bar{\nu}_{e}$ and
     $D_{s}^{\ast} {\mu}^{-} \bar{\nu}_{\mu}$ decays.
     (c)
     To reduce the theoretical uncertainties,
     much more efforts are eagerly needed to determine the shape lines
     of form factors, especially the behaviors of form factors
     at the small and middle $q^{2}$ regions.

     \section{Summary}
     \label{sec04}
     The semileptonic $\overline{B}_{s}$ ${\to}$
     $D_{s}^{({\ast})} {\ell} \bar{\nu}_{\ell}$ decays
     are induced by the weak charged current interactions
     $b$ ${\to}$ $c$ $+$ $W^{\ast}$
     ${\to}$ $c$ $+$ ${\ell}^{-}$ $+$ $\bar{\nu}_{\ell}$,
     and can provide helpful constraints to the CKM element
     $V_{cb}$ and the LFU problem highlighted in the
     semileptonic charmed $\overline{B}$ decays.
     Considering the nonfactorizable QED one-loop
     vertex corrections within SM,
     the branching ratios and ratios of branching ratios
     $R(D_{s}^{({\ast})})$ for the semileptonic $\overline{B}_{s}$
     ${\to}$ $D_{s}^{(\ast)} {\ell} \bar{\nu}_{\ell}$ decays are
     recalculated.
     It is found that
     (a) the QED effects can raise the contributions
     simultaneously from both longitudinal and transverse amplitudes,
     and consequently enhance the branching ratios according to the
     mass of the final charged lepton,
     and finally reduce the ratios $R(D_{s}^{({\ast})})$, which
     might lead to the increasing tension of the correlation
     distributions $R(D)$-$R(D^{\ast})$ between the measurements
     and SM expectation.
     (b)
     By including the QED contributions, branching ratios for the
     semileptonic $\overline{B}_{s}$ ${\to}$
     $D_{s}^{(\ast)} {\mu} \bar{\nu}_{\mu}$ decays
     are in better agreement with the available data.
     (c)
     The $SU(3)$ flavor symmetry holds basically well
     in the ratios $R(D)$-$R(D^{\ast})$ for the semileptonic charmed
     $\overline{B}_{u,d,s}$ decays.
     (d)
     Due to the strong interaction complications, the theoretical
     uncertainties of branching ratios for the semielectronic and
     semimuonic $\overline{B}_{s}$ decays predominantly come from
     the form factors.
     Besides, the precise measurements on the semitauonic
     $\overline{B}_{s}$ decays are unobtainable.
     So the verification of the nonfactorizable QED effects on
     the semileptonic $\overline{B}_{s}$ ${\to}$
     $D_{s}^{(\ast)} {\ell} \bar{\nu}_{\ell}$ decays seems to
     be impracticable for the moment.

     \section*{Acknowledgments}
     The work is supported by the National Natural Science Foundation
     of China (Grant No. 12275068),
     National Key R\&D Program of China (Grant No. 2023YFA1606000),
     and Natural Science Foundation of Henan Province
     (Grant Nos. 252300421491, 242300420250, 222300420479),
     and Science and Technology R\&D Program Joint
     Fund Project of Henan Province (Grant No. 225200810030) and
     Science and Technology Innovation Leading Talent Support
     Program of Henan Province.

     \begin{appendix}

     \section{Form factors and helicity amplitudes for the
       $\overline{B}_{s}$ ${\to}$ $D_{s} {\ell} \bar{\nu}_{\ell}$
       decays}
     \label{app01}

     We will take the conventions of
     Ref. \cite{PhysRevD.101.074513} for the $\overline{B}_{s}$
     ${\to}$ $D_{s}$ transition form factors,
     \begin{equation}
     {\langle} \, D_{s} \, {\vert} \, \bar{c} \, {\gamma}^{\mu} \, b \,
     {\vert} \, \overline{B}_{s} \, {\rangle}
     \, = \,
     f_{+}(q^{2}) \, \big[ ( p_{B_{s}}^{\mu} + p_{D_{s}}^{\mu} ) -
     \frac{ m_{B_{s}}^{2} - m_{D_{s}}^{2} }{ q^{2} } \, q^{\mu} \big]
   + f_{0}(q^{2}) \, \frac{ m_{B_{s}}^{2} - m_{D_{s}}^{2} }{ q^{2} } \, q^{\mu}
     \label{eq:ff-Bs2Ds},
     \end{equation}
     where $q$ $=$ $p_{B_{s}}$ $-$ $p_{D_{s}}$.
     The form factors $f_{+}(0)$ $=$ $f_{0}(0)$
     are generally required to cancel the singularity
     at the pole $q^{2}$ $=$ $0$.

     The helicity amplitudes $H_{\lambda}$ are expressed as,
     \begin{eqnarray}
     H_{\pm} & = & 0
     \label{eq:hp-hm-Bs2Ds}, \\
     H_{0} & = & \frac{ 2\, m_{B_{s}} \, {\vert} \vec{p} \, {\vert} }{ \sqrt{q^{2}} } \, f_{+}(q^{2})
     \label{eq:hz-Bs2Ds}, \\
     H_{t} & = & \frac{ m_{B_{s}}^{2} - m_{D_{s}}^{2} }{ \sqrt{q^{2}} } \, f_{0}(q^{2})
     \label{eq:ht-Bs2Ds}.
     \end{eqnarray}

\begin{figure}[h]
     \includegraphics[width=0.29\textwidth]{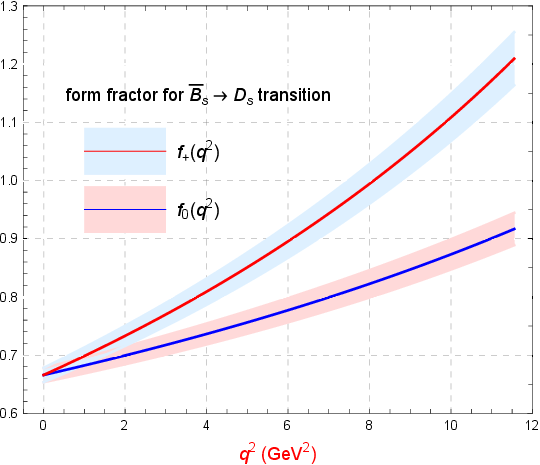} \qquad
     \includegraphics[width=0.3\textwidth]{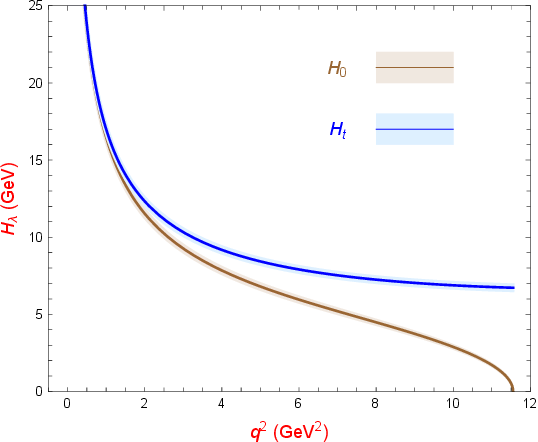}
     \caption{The shape lines of form factors (left)
        and helicity amplitudes (right) versus $q^{2}$.}
     \label{fig:ffP2P}
     \end{figure}

     Using the $z$ expansion of the Bourrely-Caprini-Lellouch (BCL)
     parametrization \cite{PhysRevD.79.013008},
     the form factors are expressed as
     (see the Appendix A of Ref. \cite{PhysRevD.101.074513}),
     \begin{eqnarray}
     f_{0}(q^{2}) & = &
     \frac{1}{1 - \displaystyle \frac{q^2}{m_{B_{c0}}^{2}} }
     \sum\limits_{n=0}^{2} a_{n}^{0} \, z^{n}(q^{2})
     \label{eq:BCL-f0}, \\
     f_{+}(q^{2}) & = &
     \frac{1}{1 - \displaystyle \frac{q^{2}}{m_{B_{c}^{\ast}}^{2}} }
     \sum\limits_{n=0}^{2} a_{n}^{+} \, \Big( z^{n}(q^{2})
    - \frac{n}{3}(-1)^{n-3} z^{3}(q^{2}) \Big)
     \label{eq:BCL-fp},
     \end{eqnarray}
     where the function $z(q^{2})$ is defined by
     \begin{equation}
     z(q^{2}) \, = \,
     \frac{ \sqrt{ t_{+}-q^{2} } - \sqrt{ t_{+} } }
          { \sqrt{ t_{+}-q^{2} } + \sqrt{ t_{+} } }
     \label{eq:BCL-zq2},
     \end{equation}
     and $t_{+}$ $=$ $( m_{B_{s}}+m_{D_{s}} )^{2}$,
     $m_{B_{c0}}$ $=$ $6.704$ GeV and
     $m_{B_{c}^{\ast}}$ $=$ $6.329$ GeV \cite{PhysRevD.101.074513}.
     With the coefficients $a_{n}^{0,+}$ listed in Table VIII of
     Ref. \cite{PhysRevD.101.074513},
     the shape lines of form factors and helicity amplitudes
     are shown in Fig. \ref{fig:ffP2P}.

     \section{Form factors and helicity amplitudes for the
       $\overline{B}_{s}$ ${\to}$ $D_{s}^{\ast} {\ell} \bar{\nu}_{\ell}$
       decays}
     \label{app02}

     We will take the conventions of
     Ref. \cite{PhysRevD.105.094506} for the $\overline{B}_{s}$
     ${\to}$ $D_{s}^{\ast}$ transition form factors,
     \begin{equation}
    {\langle} \, D_{s}^{\ast} \, {\vert} \, \bar{c} \, {\gamma}_{\mu} \, b \,
    {\vert} \, \overline{B}_{s} \, {\rangle}
     \, = \,
     \frac{ i \, 2 \, V(q^{2}) }{ m_{B_{s}} + m_{D_{s}^{\ast}} } \,
    {\varepsilon}_{{\mu}{\nu}{\rho}{\sigma}} \,
    {\epsilon}^{{\ast} \, {\nu}}_{D_{s}^{\ast}} \,
     p_{D_{s}^{\ast}}^{\rho} \,
     p_{B_{s}}^{\sigma}
     \label{eq:ff-v-Bs2Dsv},
     \end{equation}
     \begin{eqnarray}
    {\langle} \, D_{s}^{\ast} \, {\vert} \, \, \bar{c} \, {\gamma}^{\mu} \,
    {\gamma}_{5} \, b \, {\vert} \, \overline{B}_{s} \, {\rangle}
     & = &
     2 \, m_{D_{s}^{\ast}} \, A_{0}(q^{2}) \,
     \frac{ {\epsilon}^{\ast}_{D_{s}^{\ast}} \, {\cdot} \, q }{ q^{2} } \, q^{\mu}
     \nonumber \\ & + &
     ( m_{B_{s}} + m_{D_{s}^{\ast}} ) \, A_{1}(q^{2}) \,
     \Big( {\epsilon}^{{\ast} \, {\mu}}_{D_{s}^{\ast}}
    -\frac{ {\epsilon}^{\ast}_{D_{s}^{\ast}} \, {\cdot} \, q }{ q^{2} } \, q^{\mu} \Big)
     \nonumber \\ & - & A_{2}(q^{2}) \,
     \frac{ {\epsilon}^{\ast}_{D_{s}^{\ast}} \, {\cdot} \, q }{ m_{B_{s}} + m_{D_{s}^{\ast}} } \,
     \Big( p_{B_{s}}^{\mu} + p_{D_{s}^{\ast}}^{\mu} -
     \frac{ m_{B_{s}}^{2} - m_{D_{s}^{\ast}}^{2} }{ q^{2} } \, q^{\mu} \Big)
     \label{eq:ff-a-Bs2Dsv},
     \end{eqnarray}
     where $q$ $=$ $p_{B_{s}}$ $-$ $p_{D_{s}^{\ast}}$.

     The helicity amplitudes $H_{\lambda}$ are expressed as \cite{PhysRevD.105.094506},
     \begin{eqnarray}
     H_{\pm} & = & ( m_{B_{s}} + m_{D_{s}^{\ast}} ) \, A_{1}(q^{2}) \, {\mp} \,
     \frac{ 2 \, m_{B_{s}} \, {\vert} \vec{p} \, {\vert} }{ m_{B_{s}} + m_{D_{s}^{\ast}} } \, V(q^{2})
     \label{eq:hp-hm-Bs2Dsv}, \\
     2 \, m_{D_{s}^{\ast}} \, \sqrt{q^{2}} \, H_{0} & = &
     ( m_{B_{s}} + m_{D_{s}^{\ast}} ) \,
     ( m_{B_{s}}^{2} - m_{D_{s}^{\ast}}^{2} - q^{2} ) \, A_{1}(q^{2})
     - \frac{ 4 \, m_{B_{s}}^{2} \, {\vert} \vec{p} \, {\vert}^{2} }
            { m_{B_{s}} + m_{D_{s}^{\ast}} } \, A_{2}(q^{2})
     \label{eq:hz-Bs2Dsv}, \\
     \sqrt{q^{2}} \, H_{t} & = & 2 \, m_{B_{s}} \, {\vert} \vec{p} \, {\vert} \, A_{0}(q^{2})
     \label{eq:ht-Bs2Dsv}.
     \end{eqnarray}
     \begin{figure}[h]
     \includegraphics[width=0.3\textwidth]{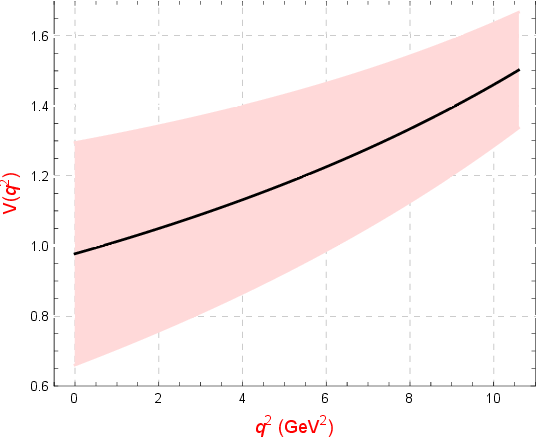} \quad
     \includegraphics[width=0.3\textwidth]{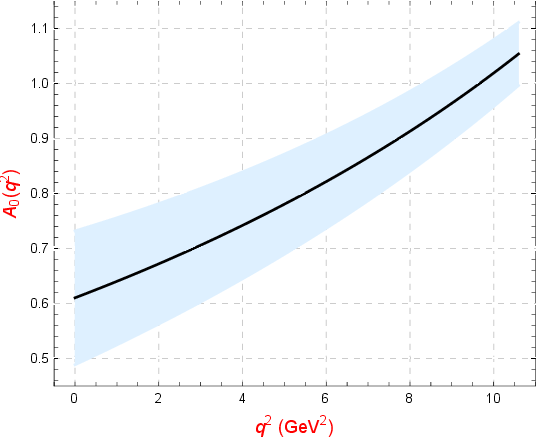} \quad
     \includegraphics[width=0.3\textwidth]{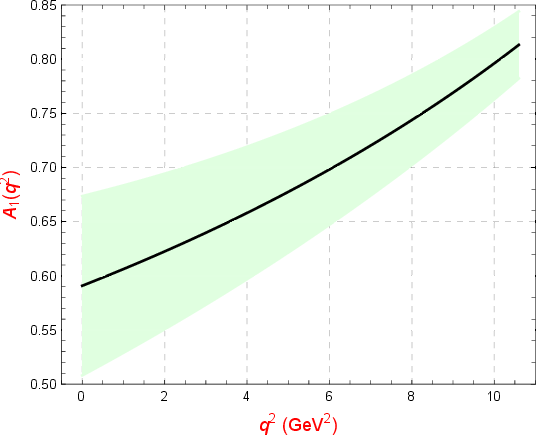} \\ ~ \\
     \includegraphics[width=0.3\textwidth]{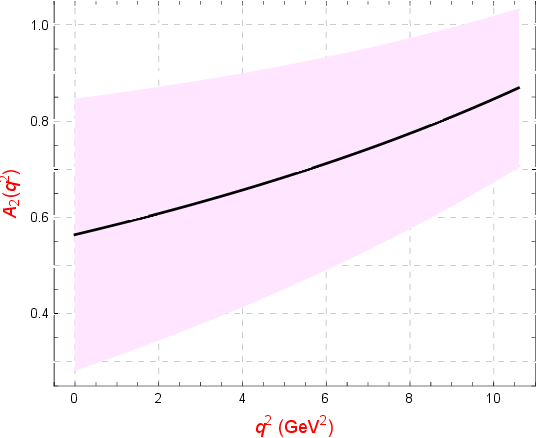} \quad
     \includegraphics[width=0.3\textwidth]{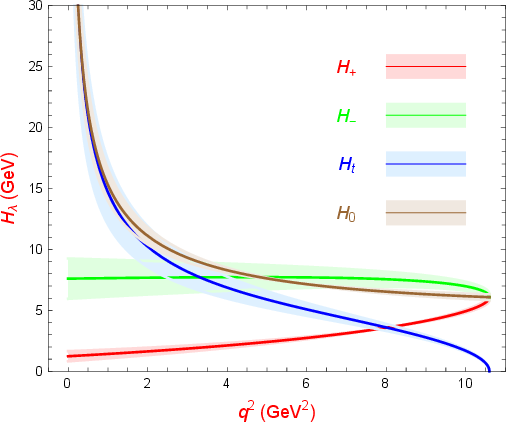}
     \caption{The shape lines of form factors and helicity amplitudes
        versus $q^{2}$.}
     \label{fig:ffP2V}
     \end{figure}

     Using the $z$ expansion of the Boyd-Grinstein-Lebed (BGL)
     parametrization \cite{PhysRevD.56.6895}, the form factors
     are expressed as \cite{PhysRevD.105.094506},
     \begin{equation}
     F_{i}(q^{2}) \, = \, \frac{1}{ P_{i}(q^{2}) } \,
     \sum\limits_{n=0}^{3} a_{n} \, z^{n}(q^{2},t_{0}),
     \qquad
     F_{i} \, = \, V \ \text{and}\ A_{0,1,2}
     \label{eq:BGL-ff},
     \end{equation}
     with the Blaschke factors $P_{i}$ embodying the pole effects and
     the poles $m_{{\rm pole},i}$ resulting from the possible particles below the
     pair production threshold $t_{+}$ with the
     $\bar{b}c$ quark content and the same quantum numbers
     as the corresponding currents,
     \begin{equation}
     P_{i}(q^{2}) \, = \, \prod\limits_{ k } z(q^{2},m_{{\rm pole},k}^{2})
     \label{eq:BGL-Blaschke-factor},
     \end{equation}
     and the variable
     \begin{equation}
     z(q^{2},t_{0}) \, = \,
     \frac{ \sqrt{ t_{+}-q^{2} } - \sqrt{ t_{+}-t_{0} } }
          { \sqrt{ t_{+}-q^{2} } + \sqrt{ t_{+}-t_{0} } }
     \label{eq:BCL-zq2},
     \end{equation}
     with $t_{+}$ $=$ $( m_{B}+m_{D^{\ast}} )^{2}$ and
     $t_{0}$ $=$ $( m_{B_{s}}-m_{D_{s}^{\ast}} )^{2}$.
     With the resonances listed in Table XII and the coefficients
     $a_{n}$ in Table XIII of Ref. \cite{PhysRevD.105.094506},
     the shape lines of form factors and helicity amplitudes
     are shown in Fig. \ref{fig:ffP2V}.

     \end{appendix}

     

     \end{document}